# Continuous monitoring of plant sub-cellular structural changes for plant and crop diseases detection by use of Intelligent Laser Speckle Classification (AI) technique

Ahmet Orun
De Montfort University, Faculty of Computing, Engineering and media. Leicester UK
Email: aorun@dmu.ac.uk. Phone: +44(0)116 3664408

## Abstract

The continuous online monitoring of early signs of plant and crop diseases, at their early stages before a potential spread, is of high importance and necessitates multi-disciplinary techniques. Within this study a proposed technique achieves this goal by exploiting laser physics, textural image analysis, and AI for *Shot hole disease*. In this technique, specific laser light with a wavelength shorter than a sub-cellular component of an inspected plant, produces an interaction within the sub-cellular components and generates laser speckle patterns which can characterize those specific plant cells' features. The generated laser speckle image data then be quantized by texture analysis and classified by Bayesian networks. Such comparative methods manage to detect the differences at sub-cellular scales, such as nuclei modification, cellular shape, or size deformation, etc. for *Shot hole* disease with high classification accuracy between the healthy and diseased plants. The technique is capable of continuous online observation and monitoring of the plant or crop diseases via a wireless network at low instrumental cost and may replace the costly ground-truth field works.
**Keywords:** *plant diseases, ground truth, laser speckle, Bayesian networks, online monitoring, image analysis*

## 1 Introduction

By the latest development in integrated technologies including online image transfer, artificial intelligence (AI), laser speckle imaging and image texture analysis, it would be possible to make remote-continuous observation of the plant and crop diseases that would be potentially used in forestry, agriculture, etc. Within the work, main fundamental method called laser speckle imaging is well established one that has been already used for decades in the areas of industrial inspection, medical applications, material science, etc. This imaging method is based on physical phenomenon so that when a rough surface is illuminated by a coherent light like laser, then the light scatters from the surface and exhibits a particular intensity distribution covering the surface with fine granular form. In earlier studies only photographic materials were used for laser speckle image formation. But with the production of digital sensor matrices (e.g. *CCD, CMOS,* etc.) the technique has become more advanced in association with digital recording facilities and for dynamic form of laser speckle analysis at higher resolution with the pixel based statistics [5]. By exploiting the digital imaging facilities, the method was later unified with different texture analysis and also learning classifiers (e.g. Bayesian) which is then called "Intelligent Laser Speckle Classification *(ILSC)*" whose earlier version was first introduced for material surface inspection in 2003 [2] and then for medical skin analysis work [1] and for medical tablet characterisation [11]. One of the most similar techniques introduced by the earlier study [12] called *Laser speckle contrast imaging* (LSCI) which is based on moving features (e.g. blood circulation in micro-veins) as is not suitable for the observations of still cellular features. LSCI method also only relies on manual image inspection made by naked eye without any use of AI method like in *ILSC*.

The proposed novel method called Intelligent laser speckle classification has potential applications for non-invasive live cells observations in real-time. The method can be used as "disease threat early warning system" for plants by continuous online monitoring of agricultural or forestry areas. On the basis of physical sciences, the study is based on a principle of laser physics so that a coherent laser light whose wavelength ($\lambda=0.65$ µm) is shorter than plant cellular features (like changes in cell size, sub-cellular component, etc.) can interact with those features and generates laser speckle patterns that characterize those specific featural changes under inspection. By the method we believe that, it would be possible to detect dynamic changes at sub-cellular scales such as genetic modification, cell size changes or cell nucleus shape deformation, etc. Such potential of use has been concluded after the ILSC experiments in which healthy and diseased leaf tissues (Figure 3) were discriminated with 87% accuracy even though those two classes of tissues looked very identical to each other under the microscope magnification (Figure 3). In this study red laser ($\lambda=0.65$ µm) is used whose wavelength is shorter than a plant cell (~60 µm) that is suitable to interact with sub-cellular features to generate speckle effects. By the technique called Laser Speckle Imaging, the speckle images are acquired by a low-cost digital camera and sampled from the image locations where laser-leaf tissue interaction is effective (generating characteristic patterns) as seen in Figure 5. In Figure 5, the diseased and healthy leaf tissues at the images of λ=6.5 μm laser Speckle sampling in 200x200 pixel windows (refers to 0.56mm$^2$ each ) with an image resolution of 2.8 μm /pixel. The pattern differences between the diseased and healthy tissues are invisible to naked eye as is also in the microscope images in Figure 4.

The instrumental part of the system includes distributed wireless image sensors located over the agricultural areas (farms, greenhouses, forests, etc.) which will transmit laser speckle image (LSI) sequences periodically over a communication network (e.g. 4G-cellular network, RFID, etc.) for close monitoring purposes [16]. The recent studies propose IoT (internet-of-things) based sensors network technology to be used in agriculture for data collection and transferring [17][18]. Optical and various sensor data transmission for agricultural monitoring is also introduced by Paul et al. (2022), as they stated that 5G mobile network can enable high-speed connection up to 20 Gigabits/sec. peak data rates for farms. The proposed ILSC system will also provide a "ground truth" information for the satellite imagery based remote sensing or Earth observation missions like Proba-V, Sentinel [20], etc. by use of above data transmission infrastructure and technologies as satellite imagery based Remote sensing mission's ground truth tasks are still major issue due to Atmospheric effects, high cost labour works, etc. [21]

## 1.2 Conventional remote sensing techniques

Traditional monitoring activities of large forest and agricultural areas have often been made by remote sensing via multispectral aerial or satellite imagery but to be necessarily supported by reliable and accurate ground truth works [26] at considerably high cost. Satellite image data in particular are the most beneficial ones for close observation of the terrestrial events that have been already used for last few decades. They are particularly becoming more and more economic with the most recent Earth observation satellite missions (e.g. Proba-V, Sentinel, etc.) as they provide high resolution spatial and spectral imagery. Several earlier studies also exploited the satellite imagery for crop and plant disease monitoring of large areas, where a comprehensive field surveys were conducted to support ground truth in one case [27] with 76 sampling plots (12 healthy and 64 stressed plants). The other study proposed a method for tracking *Mustard-Rot disease* by satellite observation via Resourcesat-1 satellite image and supported by ground measurements (16 selected plant regions and each region consisted of 5-7 subfields of 900-3600 m$^2$ ) which were also very costly [28]. As is introduced within this work, the proposed continuous ground monitoring ILSI technique may replace such high cost and long lasting labour work of field works at higher disease monitoring and ground truth accuracy.

## 1.3 Laser speckle phenomenon

According to the basic principle of "laser light-surface" interaction, when a rough surface is illuminated by a coherent light like laser then the light itself scatters from the surface exhibiting a particular intensity distribution as it looks covering the surface with a granular structure. The very fundamental formula of laser speckle image includes its pixel intensity statistics, where the standard deviation of spatial intensity variations $\sigma_s$ is equal to the mean intensity $\langle I \rangle$ for full developed (ideal) speckle pattern. This would be stated by the basic formula ; $K = \sigma_s / \langle I \rangle$ ,Where K is the speckle contrast and its value takes place between 0 and 1. If the speckle pattern is ideal then K=1. But if the speckle pattern becomes not ideal such as blurred by a diffuser or surface motion, then value of K will be shifted towards zero.

# 2 Materials and methods

### 2.1 Material sampling of leaf tissues

Shot hole disease (also called *Coryneum blight* ) is a serious fungal disease which effect plant leaves [13] generating large size holes as seen in Figure 3. The pathogen called *Wilsonomyces Carpophilos* causes the disease which ends up with falling of the leaves.

The higher the temperature the more quickly the disease is infected [14] The experimental observations have shown that at 25$^o$ environmental temperature the infection takes only 6 hours. The Fungal pathogen of Shot hole disease can persist several years in the infected plants.

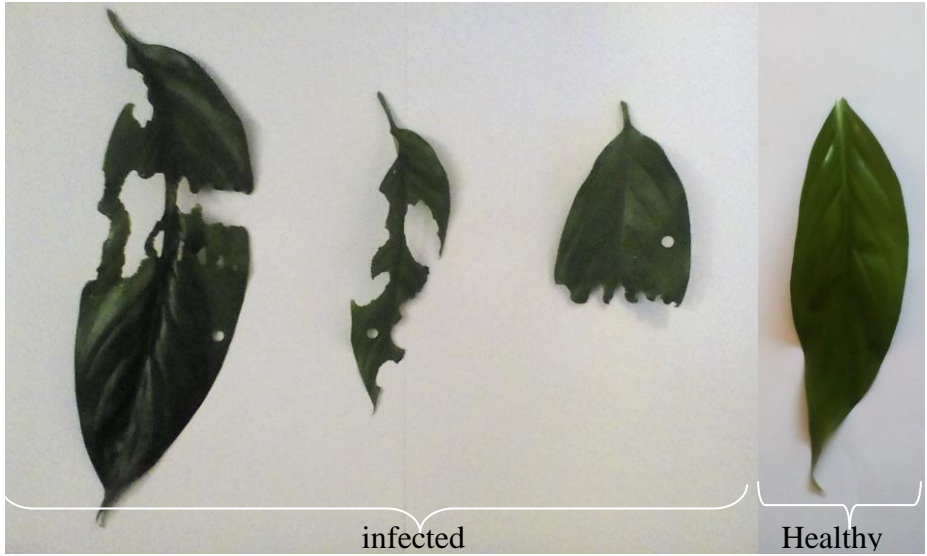

Figure 3. The samples of healthy leaf (on the right) and Infected leaves by Shot hole Disease (on the left) whose cellular or sub-cellular features are modified. (The circular holes on the leaves are because of tissue sampling for the microscopic inspection).

**2.2 Intelligent laser speckle classification**

Within this work, the novel method called Intelligent Laser Speckle Classifier (ILSC) [11] is used having been tested also in earlier studies and yielding promising results [1][2][3]. The method consists of different components such as laser speckle imaging technique [4], texture analysis [5] and Bayesian classifier [6] that are merged to provide optimum classification of laser speckle patterns each specific to different class of tissues (diseased or healthy leaf tissues)

For texture based quantization process of the LSIs, five statistical measures are used (derived from Phillips) [5]. Each texture's statistical characteristic is applied onto two different sizes of operation windows (3 x 3) which correspond to Equation. 1–4 and single measure at only 3 x 3 window size corresponds to Equation 5. Each textural measure has been applied on two 200x200 pixel areas on each LSI. As far as such texture analysis is concerned, large window size produce large edge effect at the class edges but provide more stable texture measures than small windows. In return, small window size is less stable but has smaller edge effect [7] . The texture measures used are shown by Formulas 1–5. These texture measures had been previously tested on different material surface types in the experiments and yielded satisfactory results for surface texture analysis and surface type identification. The texture values of different classes (diseased and healthy leaf tissues) have then be analysed by Bayesian classifier to differentiate the subtle microscopic characteristic patterns which are invisible to eye under microscope magnification (Figure 4) and also even in laser speckle images (Figure 5)

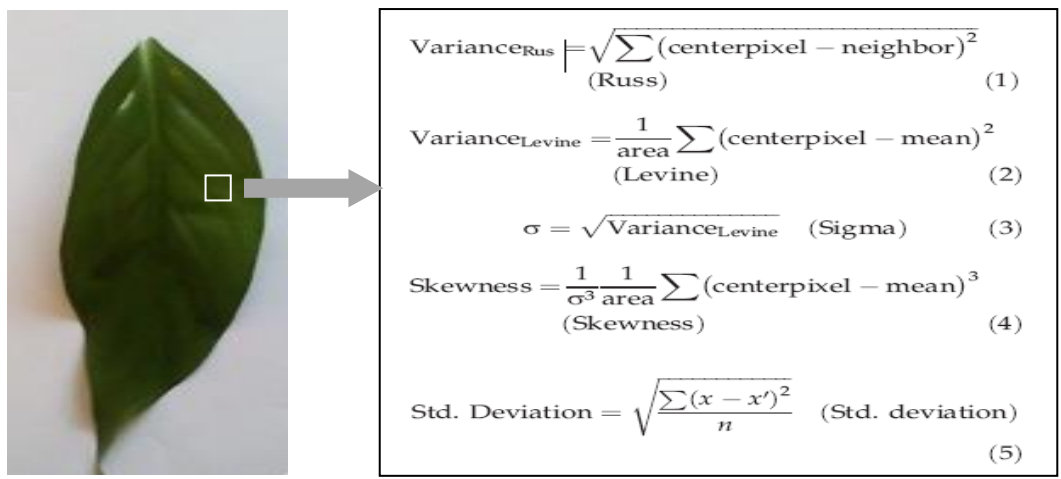

$$\text{Variance}_{Rus} = \sqrt{\sum (\text{centerpixel} - \text{neighbor})^2} \quad \text{(Russ)} \quad (1)$$

$$\text{Variance}_{Levine} = \frac{1}{\text{area}} \sum (\text{centerpixel} - \text{mean})^2 \quad \text{(Levine)} \quad (2)$$

$$\sigma = \sqrt{\text{Variance}_{Levine}} \quad \text{(Sigma)} \quad (3)$$

$$\text{Skewness} = \frac{1}{\sigma^3} \frac{1}{\text{area}} \sum (\text{centerpixel} - \text{mean})^3 \quad \text{(Skewness)} \quad (4)$$

$$\text{Std. Deviation} = \sqrt{\frac{\sum (x - x')^2}{n}} \quad \text{(Std. deviation)} \quad (5)$$

Figure 1. Texture measures applied onto texture samples taken on the laser speckle image segments where there is high interaction between the laser light and leaf tissue cellular network (for diseased and healthy leaves)

## 2.3 laser speckle image acquisition and image texture analysis

For the image acquisition, two groups of plant leaves have been selected. First group includes healthy leaf (Figure 3) and the second group includes Infected leaves by Shot hole Disease whose cellular or sub-cellular features are possibly modified. All laser speckle images as seen in figure 5 are taken by a high resolution camera at 3840x2880 pixel resolution. Diseased and healthy leaf tissues images of $\lambda=6.5$ µm laser Speckle images are sampled at 200x200 pixel (each refers to 0.56mm$^2$ area ) with an image resolution of 2.8 µm /pixel. The pattern differences in Figure 5 between the healthy and diseased image samples are invisible to naked eye. As is seen in figure 2, Laser speckle imaging setup configuration consists of the camera and a laser source by which an image of cellular tissue is acquired. The laser wavelength $\lambda= 0.65$ µm is shorter than individual cell size (~60µm) or sub-cellular components size (e.g. cell nucleus) So that the laser can interact and back scatter by conveying the characteristic pattern of cellular components to the speckle image. The surface normal makes 2º angles with laser beam and camera viewing axis. The Figure 4 also indicates that the diseased and healthy leaf tissues images taken with a microscope magnification can not be visible to naked eye under normal light illumination as well as their laser speckle image equivalences. The texture analysis is applied on each laser speckle image sample of the leaves classes (diseased/healthy) using 9 texture measures [5]. Texture image sample size are selected as 3x3 pixel windows for Area A, and 200x200 pixel windows for Area B as shown in Figure 1. It was stated that [7] large window size generates large edge effect at the class edge but generate a texture measure with higher stability than smaller window.

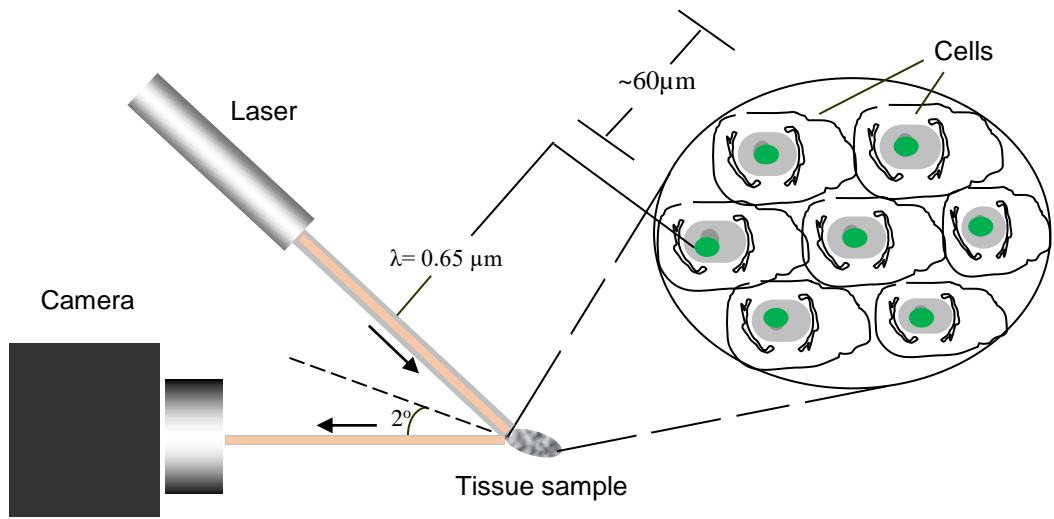

Figure 2. Laser speckle imaging setup configuration by which an image of cellular tissue is acquired. The laser wavelength $\lambda= 0.65$ µm is shorter than individual plant or crop cell size (~60µm) or sub-cellular components size (e.g. nucleus) So that the laser can interact and back scatter by conveying the characteristic pattern of cellular components to the speckle image. The surface normal makes 2º angles with laser beam and camera viewing axis.

## 2.4 AI techniques and Bayesian networks

The broad use of the artificial intelligent (AI) techniques in agricultural data analysis has become popular in last few decades such as Neural Networks (NN), Bayesian Networks (BN), Maximum Likelihood classifiers (ML) or Decision Trees (DT) in association with the contemporary high speed computational facilities, as such systems enable users to make future predictions with an adequate training (learning) process by historic data [8]. In particular Bayesian Networks have a clear advantages over its counterparts such as Neural networks by avoiding "over-fitting" phenomenon [23] and also proper graphical representation of network links and attributes [22] in terms of user friendliness and better visual comprehension. In addition to those, BN are based on logical understanding of the relational structures of the network variables [24] and they provide a flexible framework for modelling by use of limited and incompatible data [25]. Bayesian networks are well established classification method whose background information is available publicly in various sources. Bayesian networks (BN) are known in the literature as 'directed acyclic graphs' which perform knowledge representation and reasoning even under uncertainty. They are also called directed Markov fields, belief networks or causal probabilistic networks [9][10]. Its operational principles may be described by its generic Equation 6 where : P(X) is the joint probability distribution which is the product of all conditional probabilities. Pa($X_i$) is the parent set of $X_i$ (e.g. class node to decide normal/micro-collapse product) For the experiments the BN utility called PowerPredictor$^{TM}$ is used. The utility is used for automated network construction which accepts the continuous variables and uses Markov conditions to obtain a collection of CI statements from the network [14]. All valid CI relations can also be extracted from the topology of the network. The algorithm examines information of two related variables from a data set and decides if two variables are dependent. It also examines how close the relationship is between those variables. This information is called conditional mutual information of two variables Xi, Xj which may be denoted as in Equation 7.

$$P(X) = \prod_i P(X_i | pa(X_i)) \qquad (6)$$

$$I(X_i, X_j | C) = \sum_{x_i, x_j, c} P(x_i, x_j, c) \log \frac{P(x_i, x_j | c)}{P(x_i | c) P(x_j | c)} \qquad (7)$$

In Equation 7, C is a set of nodes and c is a vector (one instantiation of variables in C). If I(Xi, Xj |C) is smaller than a certain threshold t, then we can say Xi and Xj are conditionally independent. This criteria is the basis of an automated network construction with the links between the network variables (nodes) as shown in Figure 6. In the following experimental stages, the laser speckle image sampling is followed by the textural data set construction then the classification method called Bayesian networks is used to distinguish between the diseased and healthy plant tissue whose micro scale changes are not visible to normal eye.

## 3 Results and discussion

Within this work, the experimental tests have proven that fungal pathogen population has an impact on the leaf tissue at the homogenous form of cellular network level, generating a textural structural effect which can only be measurable by specific LSI textural analyses [5] as seen in Figure 5 and also in association with AI methods.

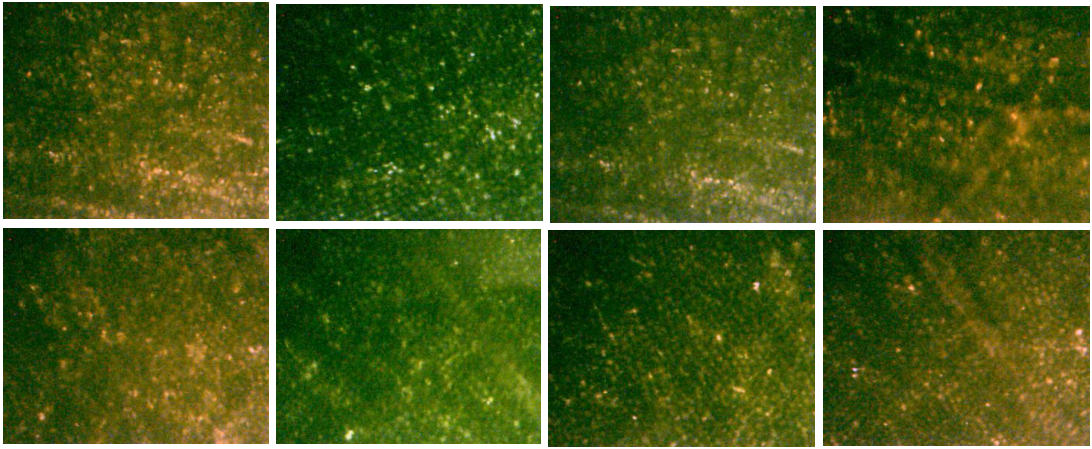

Figure 4. Diseased (top) and healthy (bottom) leaf tissues images with a microscope magnification of 150x.
As is seen the characteristic differences between diseased and healthy tissues can not be visible to human eye under normal light illumination.

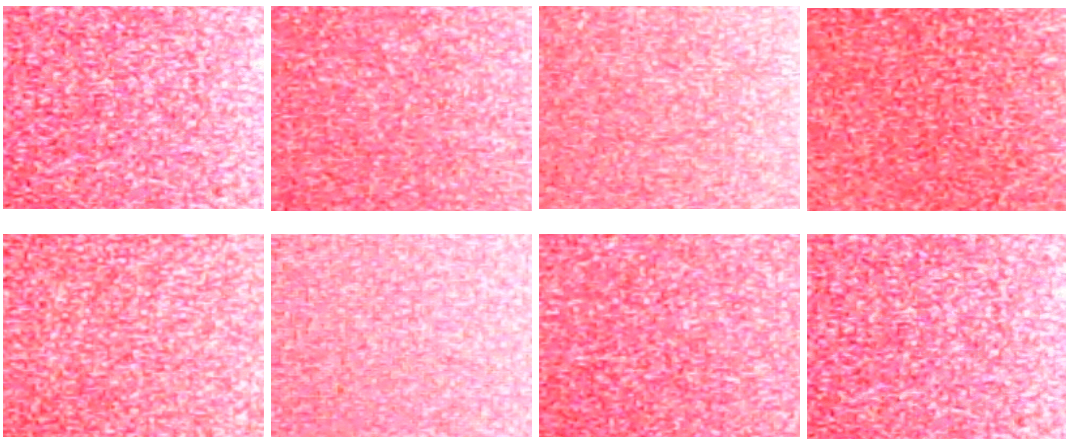

Figure 5. The typical Diseased (top) and healthy (bottom) leaf tissues images of λ=6.5 µm laser Speckle sampling at 200x200 pixel (0.56mm$^2$) with an image resolution of 2.8 µm /pixel. The pattern differences between the top and bottom rows are invisible to naked eye and can only be classified by statistical LSI technique.

## 3.1 Experimental data set

The data set consists of 40 cases which correspond to leaf samplings taken from almost equal classes of healthy and diseased leaf populations. The attributes of data set belong to 9 texture measures whose pixel based statistics are already explained in Figure 1. The reduced form of *40x10* data set is shown in Table 1. The first group of texture measures (index1) refer to interior laser speckle band region which is just around the edge of laser bright spot area (marked as A), and the second group refers to exterior laser speckle band region (marked as B) as shown in Figure1. The laser speckle image data sampling is selected with an optimum pixel window as each window has to cover enough texture primitives for the pixel based textural statistics. Whereas the extremely larger pixel windows increase the computational cost unnecessarily. The sampling has also to be done to justify homogeneity of the texture windows which is inevitable for an accurate texture analysis [3]

Table I. Laser speckle image texture values of the leaf samplings. The first group of texture measures (index1) refer to interior laser speckle band region, and the second group refers to exterior laser speckle band region as shown in Figure1. Classes "h" and "d" refer to healthy and diseased plants respectively (here *40x10* data set is reduced for the display purpose)

| sample No. | russ1 | levine1 | sigm1 | skew1 | russ2 | levine2 | sigm2 | skew2 | stdev | Class |
|---|---|---|---|---|---|---|---|---|---|---|
| 1 | 421 | 111 | 10.54 | 1.78 | 91 | 169 | 13 | -1.43 | 30.73 | h |
| 2 | 524 | 190 | 13.78 | -0.2 | 318 | 258 | 16.06 | -2.65 | 32.8 | h |
| 3 | 121 | 11 | 3.32 | 3.06 | 22 | 13 | 3.61 | 0.27 | 16.93 | h |
| 4 | 219 | 26 | 5.1 | 2.24 | 28 | 38 | 6.16 | -0.02 | 20.58 | h |
| 5 | 75 | 13 | 3.61 | 1.51 | 80 | 23 | 4.8 | 0.37 | 10.08 | h |
| 6 | 281 | 63 | 7.94 | 0.97 | 83 | 101 | 10.05 | -0.82 | 20.96 | h |
| . | 428 | 753 | 27.44 | -1.52 | 304 | 862 | 29.36 | -2.05 | 40.57 | d |
| . | 209 | 48 | 6.93 | 0.63 | 62 | 75 | 8.66 | -1.09 | 28.64 | d |
| . | 177 | 25 | 5 | 2.74 | 16 | 37 | 6.08 | 0.37 | 23.29 | d |
| . | 163 | 19 | 4.36 | 2.76 | 44 | 26 | 5.1 | 0.42 | 20.54 | d |
| 40 | 163 | 25 | 5 | 2.34 | 72 | 37 | 6.08 | 0.28 | 22.66 | d |

## 3.2 Classification by Bayesian Networks

By the proposed method called Intelligent laser speckle classification in which the appropriate imaging instruments (low-cost camera and low level (1mW) red laser source are used as well as 9 texture measures shown in Figure1, it is possible to distinguish between the diseased and healthy leaf tissue groups by use of Bayesian classifier PowerPredictor™ (Figure 6) with 87% accuracy. The test results strengthen the idea that the laser beam at $\lambda=6.5$ micron wavelength can interact with relatively larger cellular or sub-cellular features such as the possible changes in cell membrane, cell nucleus or even nucleus content, and also unveil the changes between the two group of tissues like before and after chemical treatment or diseased and healthy situations.

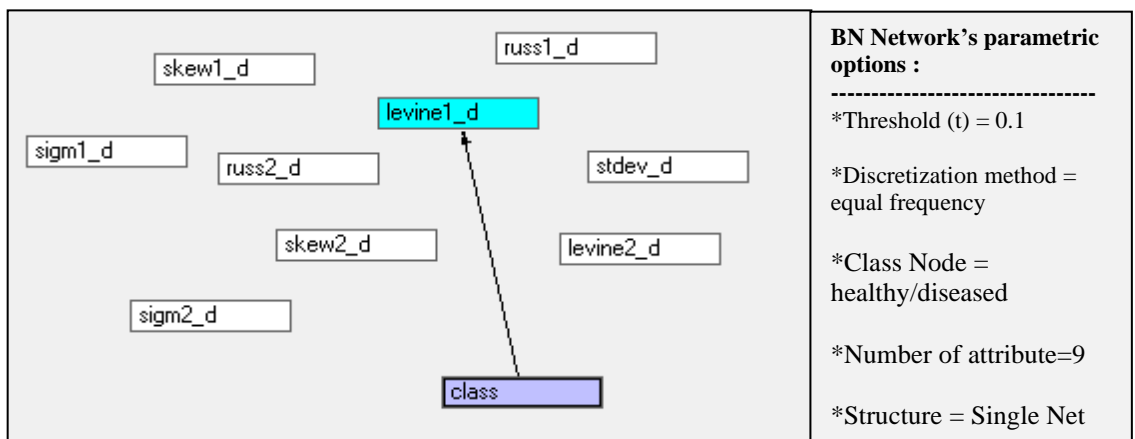

**BN Network's parametric options :**
--------------------------------
*Threshold (t) = 0.1

*Discretization method = equal frequency

*Class Node = healthy/diseased

*Number of attribute=9

*Structure = Single Net

Figure 4. Attributes network build after BN training by whom max 87% Classification accuracy is obtained to distinguish between the modified and normal sub-cellular contents of two classes. The same attributes are also exhibited in Table 1. (the annotations "d" refer to discretisation for each attribute node)

As is seen in BN network (Figure 6) the cellular or sub-cellular changes before and after the side effect of Shot hole disease could be classified through the texture measure Levine1 whose image sampling scan is achieved by (3x3) pixel window. The rest of texture measures are not linked since they would make less or equal contribution to the classification results. In the Bayesian network setup, the optimum system parameters are selected for the best output such as; node connection *threshold* (t) = 0.1 is selected to include highest number of attributes into the network (as described in Equation 7), The discretization method *Equal frequency* is also selected to obtain maximum classification accuracy with the available data structure. We have to note that, the nodes those are not connected are not excluded from the classification process but their inclusion to the network would not increase the classification accuracy. The statistical results such as Confusion Matrix and ROC indices are also shown in Table II which are generated automatically by PowerPredictor$^{TM}$

Table II. Results of classification for Test 1 and 2 with different network parameters selections

| **Confusion Matrix (Test 1)** | | | | **Confusion Matrix (Test 2)** | | | |
|---|---|---|---|---|---|---|---|
| **Predicted** | diseased | Healthy | Lift Index | **Predicted** | diseased | Healthy | Lift Index |
| diseased | 0000008 | 0000002 | 0.66 | diseased | 0000006 | 0.0 | 0.80208 |
| healthy | 0000001 | 0000009 | 0.69 | healthy | 0000002 | 0000008 | 0.7 |
| **Classification accuracy : 85%** | | | | **Classification accuracy : 87.5%** | | | |
| **ROC Indices** | | | | **ROC Indices** | | | |
| diseased | 0.0 | 0.82 | | diseased | 0.0 | 0.95 | |
| healthy | 0.77 | 0.0 | | healthy | 0.93 | 0.0 | |

## 4. Proposed LSI system configuration

The proposed plants or crop fields observation system (Figure 7) consists of LSI device units network which transmit continuous LSI data wirelessly to the Ground Control Station (GCS) to be used in association with satellite image data of the corresponding agricultural fields. Each LSI device obtain regular LSI image sampling (e.g. daily, weekly, etc.) of the plants or crops under the closed observation. The LSI device units are distributed over the test field in an optimum way depending on the specification of the plant or crop type. The operational principle of each unit is already described by detailed configuration in Figure 2. The Ground control station implement data gathering and data fusion for the targeted disease monitoring and also early warning function to prevent the disease.

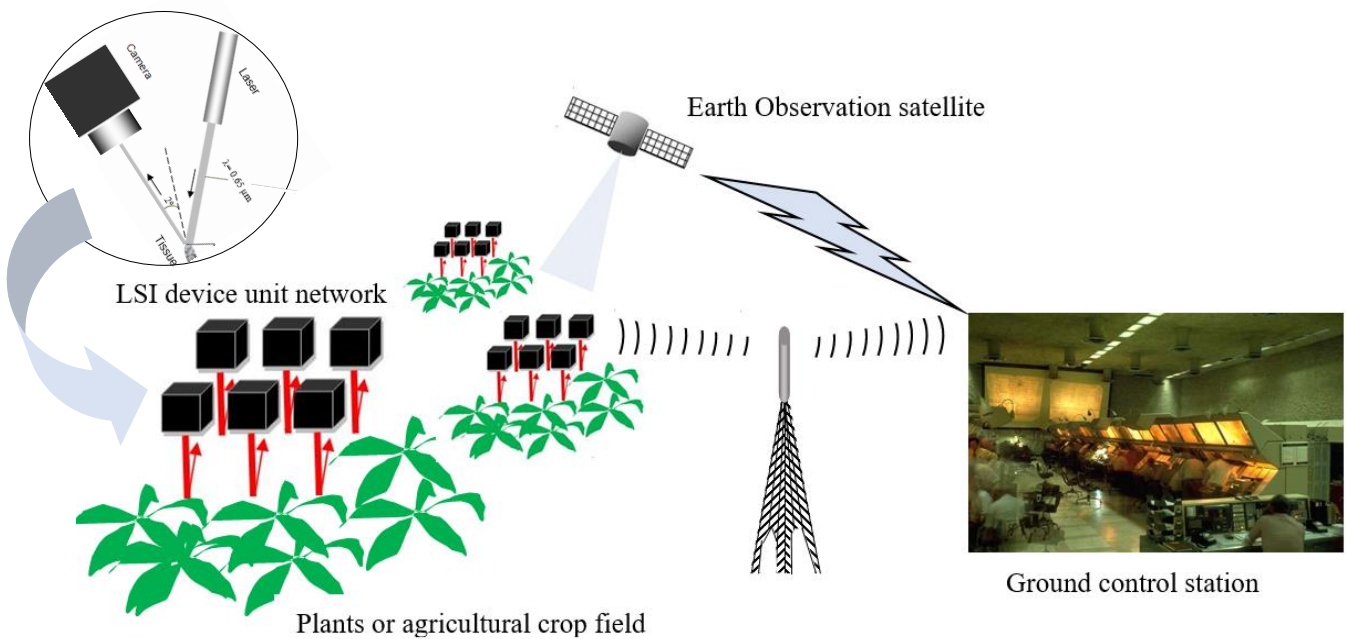

Figure 7. The system configuration for LSI image data collection, fusion and analysis for the diseases monitoring can be implemented at very low cost due to its instrumental simplicity (each system unit consists of simple CMOS camera and low-power laser source). The LSI device unit matrices are distributed optimally over the plant or agricultural crop fields. The operational principle of each unit refers to detailed configuration as described in Figure 2. The Ground control stations undertake data fusion and diseases monitoring-early warning tasks.

# Conclusion

By use of Intelligent laser speckle classification (*ILSC*) technique in association with the Bayesian Networks (PowerPredictor™ utility) a classification result with max 87.5% accuracy has been obtained to distinguish between the healthy and diseased plant or crop cells exposed to laser speckle effect. The test results have proven the theory that the laser beam at λ=6.5 micron wavelength can interact with plant's cellular or sub-cellular features such as the cell membrane, cell nucleus or sub-nucleus content. The method is also promising to unveil the subtle changes between the two group of tissues like before and after a chemical treatment, diseased or healthy conditions, low or high product quality, etc. whatever factor effective at the cellular scale. The method seems to be quite robust due to its comparative (qualitative) nature. It has also a potential to be utilized as plant and crop monitoring system applicable to various areas such as forestry, agriculture, botanic gardens, etc. and also to be used as disease early warning system to take prompt action before its spread further. The proposed monitoring ILSI technique may also replace such high cost and long lasting labour work of agricultural ground-truth field works at higher monitoring frequency and higher ground truth accuracy.